\documentclass[pdflatex,sn-mathphys-num]{sn-jnl}

\usepackage{graphicx}        %
\usepackage{multirow}        %
\usepackage{amsmath,amssymb,amsfonts} %
\usepackage{amsthm}          %
\usepackage{mathrsfs}        %
\usepackage[title]{appendix} %
\usepackage{xcolor}          %
\usepackage{textcomp}        %
\usepackage{manyfoot}        %
\usepackage{booktabs}        %
\usepackage{algorithm}       %
\usepackage{algorithmicx}    %
\usepackage{algpseudocode}   %
\usepackage{listings}        %
\usepackage{subcaption}      %
\usepackage{caption}         %
\usepackage{physics}         %
\usepackage{bm}              

\theoremstyle{thmstyleone}%

\theoremstyle{thmstyletwo}%

\theoremstyle{thmstylethree}%
\begin{document}

\title[Article Title]{Hybrid Vision Transformer and Quantum Convolutional Neural Network for Image Classification}


\author[1,2]{\fnm{Mingzhu} \sur{Wang}}

\author*[1,3]{\fnm{Yun} \sur{Shang}}\email{shangyun@amss.ac.cn}

\affil[1]{\orgdiv{Academy of Mathematics and Systems Science}, \orgname{Chinese Academy of Sciences},  \city{Beijing}, \postcode{100190}, \country{China}}
\affil[2]{School of Mathematical Sciences, University of Chinese Academy of Sciences, Beijing, 100049, China}
\affil[3]{State Key Laboratory of Mathematical Science, Academy of Mathematics and Systems Science, Chinese Academy of Sciences, Beijing, 100190, China}


\abstract{Quantum machine learning (QML) holds promise for computational advantage, yet progress on real-world tasks is hindered by classical preprocessing and noisy devices. We introduce ViT-QCNN-FT, a hybrid framework that integrates a fine-tuned Vision Transformer with a quantum convolutional neural network (QCNN) to compress high-dimensional images into features suited for noisy intermediate-scale quantum (NISQ) devices. By systematically probing entanglement, we show that ansatzes with uniformly distributed entanglement entropy consistently deliver superior non-local feature fusion and state-of-the-art accuracy (99.77\% on CIFAR-10). Surprisingly, quantum noise emerges as a double-edged factor: in some cases, it enhances accuracy (+2.71\% under amplitude damping). Strikingly, substituting the QCNN with classical counterparts of equal parameter count leads to a dramatic 29.36\% drop, providing unambiguous evidence of quantum advantage. Our study establishes a principled pathway for co-designing classical and quantum architectures, pointing toward practical QML capable of tackling complex, high-dimensional learning tasks. }




\maketitle

\section{Introduction}\label{sec1}
Quantum information science is rapidly advancing, with quantum machine learning (QML) emerging as a particularly promising frontier that explores the potential advantages of quantum computation for complex data analysis tasks \cite{Bharti22NISQ, Cerezo21VQA, biamonte2017quantum, huang2022quantum, intelligence2023seeking}. Leveraging quantum phenomena such as superposition and entanglement, QML enables novel computational paradigms \cite{dunjko2018machine, lloyd2020quantum}. Within QML, quantum neural networks (QNNs), implemented as parametrized quantum circuits (PQC) \cite{Benedetti19PQC, mitarai2018quantum, schuld2019quantum}, can surpass classical models for specific problems \cite{abbas2021power, caro2021encoding}, particularly in hybrid quantum-classical schemes that exploit the strengths of classical deep learning while accommodating the constraints of current noisy intermediate-scale quantum (NISQ) devices \cite{Preskill18NISQ, schuld2021machine, PerdomoOrtiz18Opportunities, cerezo2021variational, liao2024machine}.

Among QNN architectures, quantum convolutional neural networks (QCNNs) \cite{Cong19QCNN, beer2020training} have attracted attention for image classification, drawing inspiration from classical CNNs. Fully quantum QCNNs process all layers using qubits and quantum operations \cite{zheng2023design, hur2022quantum}, resembling the multiscale entanglement renormalization ansatz \cite{cong2019quantum, vidal2007classical}. While QCNNs have achieved high accuracy on simple datasets such as MNIST and Fashion-MNIST, their applicability to color images and more complex, high-dimensional tasks remains limited \cite{hur2022quantum}, largely due to reliance on basic feature extraction, which may fail to capture intricate correlations present in real-world data \cite{ Schuld21Kernel, ruiz2025quantum}.

In parallel, Vision Transformers (ViTs) \cite{DosoViTskiy21ViT, touvron2021training} have demonstrated exceptional capacity for hierarchical feature extraction via self-attention \cite{yosinski2014transferable, zhou2024vision, zhou2022generalized}, and have been applied in diverse domains including multimodal fusion \cite{wang2025hybrid}, high-fidelity image matting \cite{yao2024vitmatte}, and medical diagnosis \cite{das2025novel}. However, their computational cost scales quadratically with input size, motivating hybrid quantum-classical strategies that offload feature compression to classical models while exploiting QCNNs for high-order, non-local correlations. The inherent entanglement in QCNNs ($S_{\text{VN}}>0$) allows them to encode complex feature interactions that would otherwise require extremely deep classical architectures \cite{west2023towards}.

Combining these insights, we propose ViT-QCNN-FT, a hybrid framework that integrates a fine-tuned ViT with a QCNN, enabling efficient compression of high-dimensional images into feature representations suitable for NISQ devices. Systematic experiments demonstrate the impact of quantum encoding methods, QCNN ansatzes, and quantum noise on model performance. The superiority of the model is verified through ablation studies. Replacing the QCNN with a classical CNN of comparable parameter count highlighted the efficiency of the QCNN. Furthermore, analysis of the entanglement entropy distribution reveals that QCNN ansatzes exhibit progressively enhanced entangling capability across layers, facilitating non-local feature fusion. Convolution ansatzes with more uniformly distributed entanglement entropy achieve better performance and greater robustness to noise. Additionally, we observe the dual nature of quantum noise, suggesting that it could be harnessed as a potential resource. The results under 18 QCNN ansatzes demonstrate the significance of optimizing quantum circuit ansatzes.

\begin{figure}[p]
\centering
\includegraphics[width=1\linewidth]{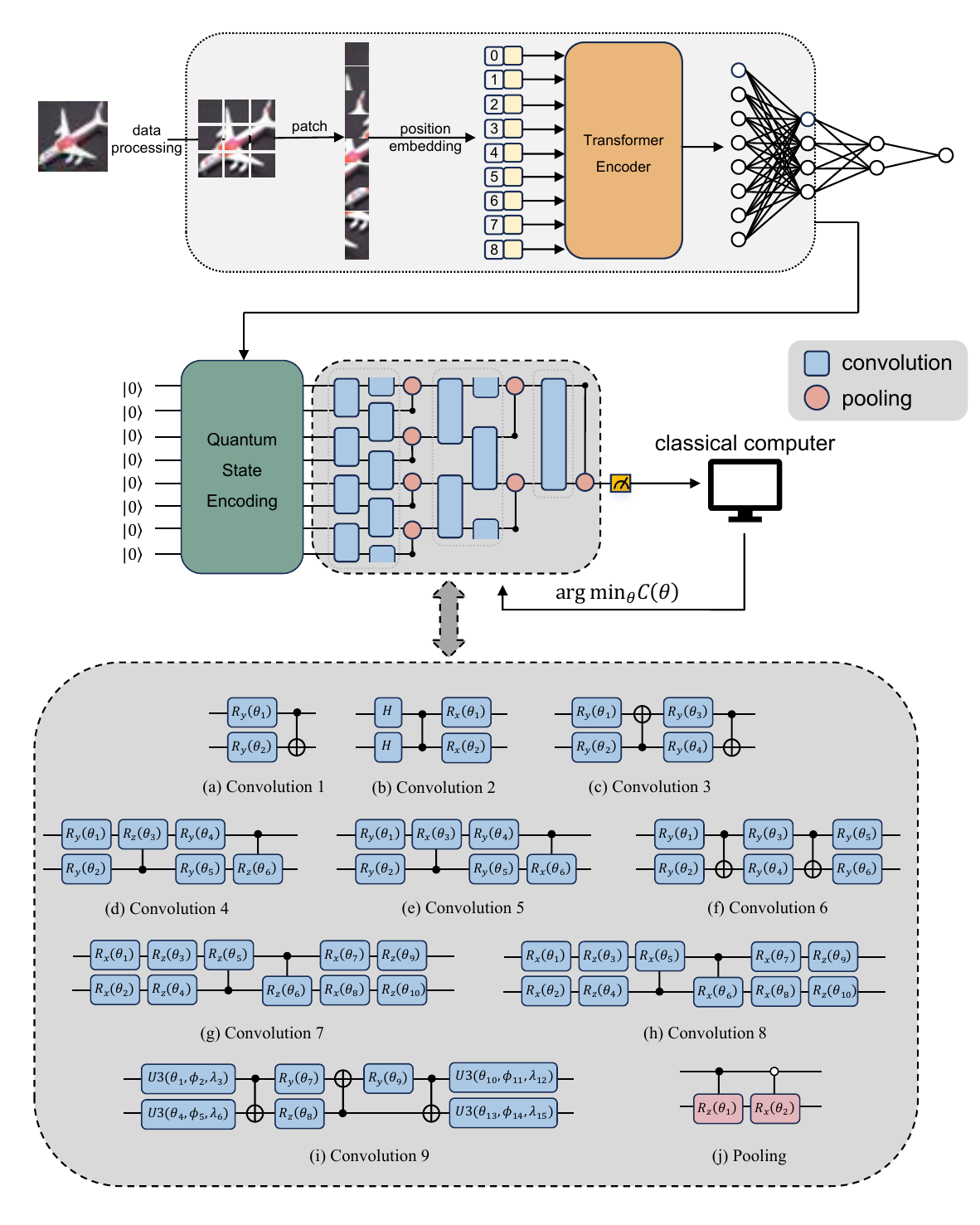}
\caption{\label{ViT-QCNN-FT} Overall model diagram of ViT-QCNN-FT and parameterized quantum circuits utilized in the convolutional and pooling layers. The pre-trained ViT is fine-tuned to act as a feature extractor. The extracted features are then encoded into quantum states (green block), followed by a QCNN performing the classification task. The QCNN consists of two primary components: convolutional filters (blue blocks) and pooling operations (red circles). $R_{\sigma}(\theta)$ represents a rotation gate around the $\sigma$-axis of the Bloch sphere by an angle $\theta$, while $H$ denotes the Hadamard gate. $U3(\theta,\phi,\lambda)$ is a general single-qubit gate, which can be expressed as $U3(\theta,\phi,\lambda)=R_z(\phi)R_z(-\pi/2)R_z(\theta)R_z(\pi/2)R_z(\lambda)$.}
\end{figure}

\section{Results}\label{sec2}
The overall algorithm is illustrated in Fig.\ref{ViT-QCNN-FT}. The pre-trained ViT is fine-tuned for feature extraction. Subsequently, classical data are encoded into quantum states, and finally, a QCNN is utilized for classification. 

We conducted the simulations in four parts using 18 QCNN ansatzes (see Section \ref{sec332}): 
(1) Encoding comparison: We compared ViT-QCNN-FT under three different quantum state encoding methods. Quantum encoding significantly influences model performance. We find amplitude encoding performs the best, and compressing the data to 10\% is enough. 
(2) Noise robustness: We simulated ViT-QCNN-FT with amplitude encoding under four types of quantum noise at intensities 0.01 and 0.05. Quantum noise can be beneficial in some cases.
(3) Feature extractor ablation: We simulated QCNN with other feature extractors and found that the average accuracy decreased by 6.64\%-40.56\%. This demonstrates the effectiveness of fine-tuned ViT. 
(4) Quantum efficiency: To evaluate the efficiency of QCNN in ViT-QCNN-FT, we replaced the quantum component of ViT-QCNN-FT with classical CNNs that have an equal or greater number of parameters. For the same parameter number, the average accuracy difference of 29.36\% demonstrates the quantum efficiency. 

\subsection{Encoding comparison}\label{subsec21}
\begin{figure}[h]
\centering
\includegraphics[width=1\linewidth]{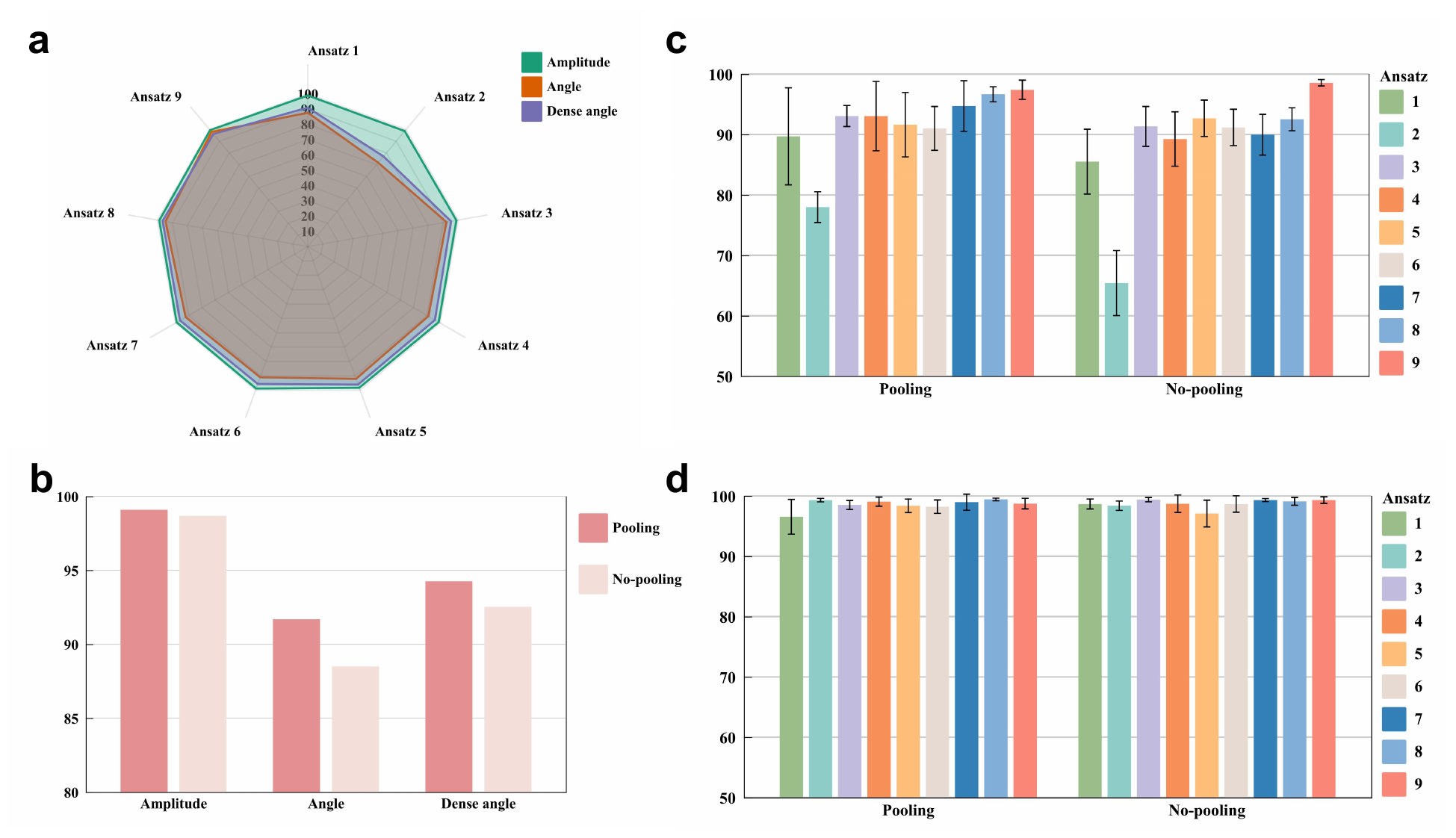}
\caption{\label{QuantumEncoding} \textbf{a}, The average accuracies of 9 ansatzes under three quantum state encoding methods (10 qubits). \textbf{b}, Comparison of pooling ansatzes and no-pooling ansatzes under three quantum state encoding methods (10 qubits). \textbf{c}, The results of 18 ansatzes under the angle encoding (10 qubits). \textbf{d}, The results of 18 ansatzes under amplitude encoding (8 qubits). }
\end{figure}

The comparison results of the three encoding methods are shown in Fig.\ref{QuantumEncoding}, which reveals three key results: 
(1) Encoding comparison: amplitude encoding consistently surpasses both angle and dense angle encodings in terms of accuracy and stability, particularly in some ansatzes (e.g., ansatzes 1 and 2). The results from angle encoding and dense angle encoding suggest that during the classical feature extraction, excessive compression of classical data may result in the loss of significant classification information. Moreover, if the data is overly compressed, a more complex and parameter-rich quantum circuit may be required. 
(2) Pooling effect: pooling and no-pooling ansatzes demonstrate a small difference in overall performance. The classical pooling operation is essentially an information compression mechanism. In a quantum system, this concept manifests as a trace operation on a specific part of the quantum system, corresponding to ending the quantum operation on certain qubits in the quantum circuit. Therefore, for QCNN, the design focus may not necessarily be on the choice of pooling ansatz, but rather on the translationally invariant design of the convolution layer and the trace operation immediately following it. 
(3) Ansatzes impact: even under the same encoding, there are performance variations across ansatzes. As shown in Fig.\ref{QuantumEncoding}\textbf{c}, under angle encoding, the accuracy rates of the best ansatz and the worst ansatz can even differ by as much as 33.15\%. The optimal ansatz is 9 no-pooling. Therefore, the fact that the average performance of the pooling ansatzes is slightly better than that of the no-pooling ansatzes does not mean that the no-pooling ansatzes will not be the optimal ansatz. 
The choice of quantum state encoding and ansatz has a significant impact on the ViT-QCNN-FT results and is task-specific. 

When reducing the qubit count from 10 to 8 in the amplitude-encoded ViT-QCNN-FT, the number of classical features that can be encoded decreases from 1024 to 256. Despite this reduction, as shown in Fig.\ref{QuantumEncoding}\textbf{d}, we achieve comparable performance. The ViT-QCNN-FT attains the highest accuracy of 99.53\% under ansatz 3 no-pooling, matching the performance of the 10-qubit amplitude encoding. This demonstrates that for binary classification on the CIFAR-10 dataset, using ViT-QCNN-FT, compressing the 3072-dimensional image data to approximately 10\% of its original dimensionality within the classical feature extractor is adequate for subsequent quantum operations. Therefore, we employ 8-qubit amplitude encoding in the subsequent experiments. 

To further validate the efficacy of ViT-QCNN-FT, we extend experiments to complete CIFAR-10 classes using 8-qubit amplitude encoding, as illustrated in Supplementary Tables 2 and 3 (Supplementary Information). For the systematic ablation studies on quantum noise and ansatz selection that follow, we focus on the binary task (classes 0 and 1) to enable a controlled comparison.

In summary, amplitude encoding yields the best performance among the methods tested, while the inclusion of a pooling ansatz has a negligible impact. The optimal ansatz is found to be highly dependent on the specific experimental configuration.

\subsection{Noise robustness}\label{subsec22}
In the current NISQ era of quantum computing, quantum noise may impact the performance of quantum machine learning models. To understand how various types and intensities of quantum noise affect ViT-QCNN-FT, we introduced four common types of quantum noise into our experiments: bit flip noise, phase flip noise, amplitude damping noise, and depolarization noise, with noise intensities set at 0.01 and 0.05, respectively. The quantum noise was added after the convolutional and pooling layers of the QCNN. In the no-pooling QCNN ansatzes, the noise was only added after the convolutional layer. 

Bit flip noise models classical bit errors in quantum systems, flipping $\ket{0} \leftrightarrow \ket{1}$ with probability $p$:
\begin{equation}
\mathcal{E}_{\mathrm{BF}}(\rho) = (1-p)\rho + p\sigma_x\rho\sigma_x. 
\end{equation}
Phase flip noise destroys quantum coherence by adding $\pi$-phase shift to $\ket{1}$ with probability $p$: 
\begin{equation}
\mathcal{E}_{\mathrm{PF}}(\rho) = (1-p)\rho + p\sigma_z\rho\sigma_z. 
\end{equation}
Amplitude damping noise simulates energy dissipation ($\ket{1} \rightarrow \ket{0}$) with decay probability $p$: 
\begin{equation}
\mathcal{E}_{\mathrm{AD}}(\rho) = \sum_{k=0}^{1} K_k\rho K_k^\dagger,\quad
K_0 = \begin{bmatrix} 1 & 0 \\ 0 & \sqrt{1-p} \end{bmatrix},\ 
K_1 = \begin{bmatrix} 0 & \sqrt{p} \\ 0 & 0 \end{bmatrix}
\end{equation}
Depolarizing noise induces complete decoherence with error probability $p$:
\begin{equation}
\mathcal{E}_{\mathrm{Depol}}(\rho) = (1-p)\rho + \frac{p}{3}\sum_{i=x,y,z}\sigma_i\rho\sigma_i
\end{equation}

\begin{figure}[h]
\centering
\includegraphics[width=1\linewidth]{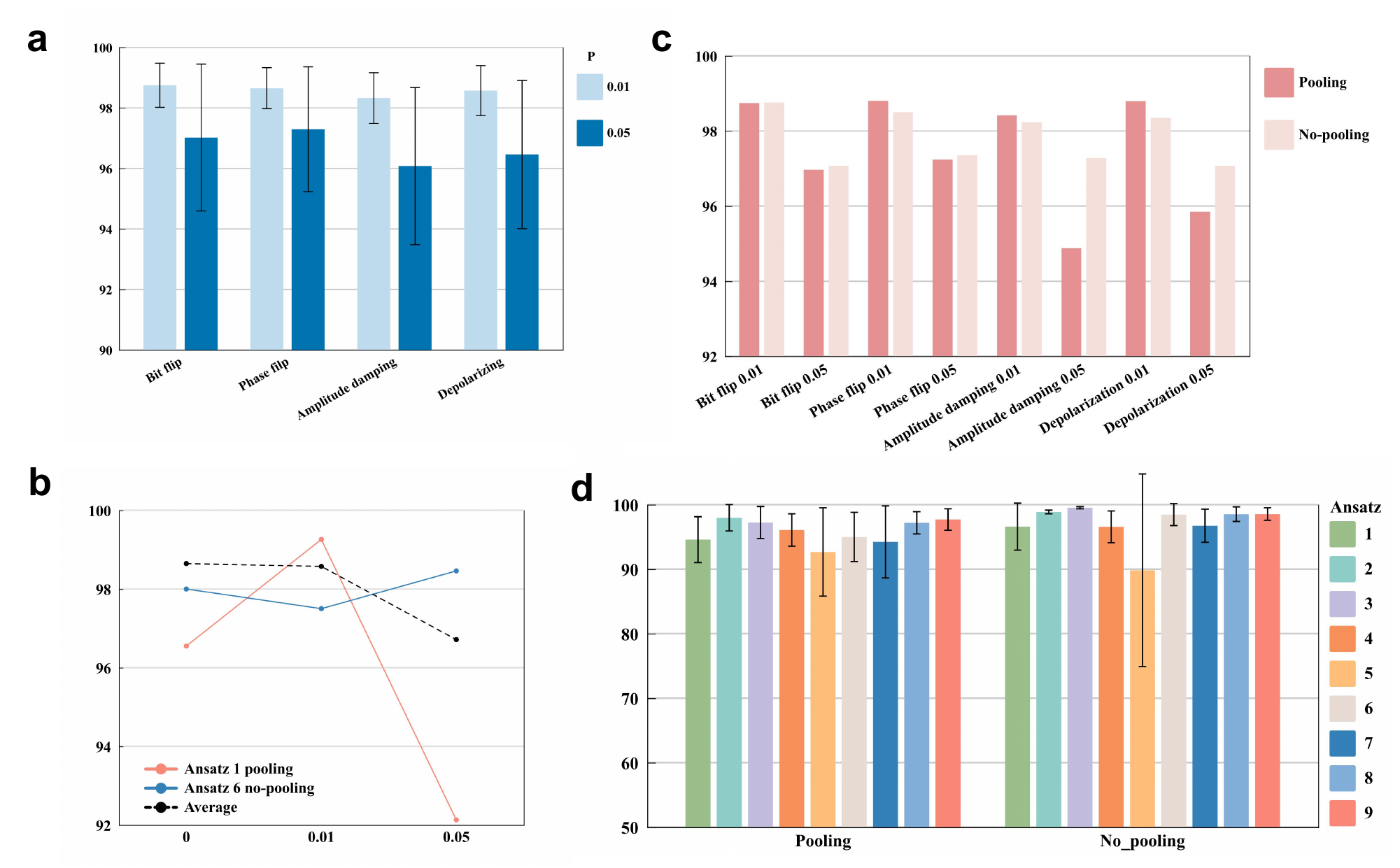}
\caption{ \label{QuantumNoise}\textbf{a}, The influence of noise intensity on the results under different types of quantum noise. \textbf{b}, The effects of pooling and no-pooling under different noise types and intensities. \textbf{c}, The average accuracy of ansaztes under different noise intensities. \textbf{d}, The results of 18 ansatzes under the depolarization noise intensity of 0.05. }
\end{figure}

Fig.~\ref{QuantumNoise} presents the results of ViT-QCNN-FT with quantum noise. The complete experimental results can be found in Supplementary Tables 4 and 5 (Supplementary Information). Overall, the model demonstrates strong noise robustness. Three critical observations emerge: 
(1)Noise's two sides: as shown in Fig.\ref{QuantumNoise}\textbf{a}, with the noise intensity increasing, the performance of the model gradually declines, but it still maintains a relatively high recognition accuracy. Interestingly, in some cases, noise does not always correlate with decreased accuracy. For example, as shown in Fig.\ref{QuantumNoise}\textbf{b}, under amplitude damping noise intensity 0.01, the average accuracy of ansatz 1 pooling increases by 2.71\% compared to the noiseless condition. Under depolarization noise intensity 0.05, the average accuracy of ansatz 6 no-pooling even surpasses the noiseless condition. Moderate noise introduces perturbations that may compel models to learn robust features, consequently improving generalization in noisy environments. However, the accuracy and stability of ViT-QCNN-FT using ansatz 5 no-pooling decrease significantly when the depolarization noise intensity increased from 0.01 to 0.05 (Fig.\ref{QuantumNoise}\textbf{d}). This indicates that this particular ansatz is unsuitable for the task when the depolarization noise is amplified. 
(2) Pooling effect: we found that when the noise intensity was low, pooling ansatzes performed better than non-pooling ansatzes, but when the noise increased, the opposite was true(Fig.\ref{QuantumNoise}\textbf{c}). 
(3) Ansatzes impact: Fig.\ref{QuantumNoise}\textbf{d} shows the results under a depolarization noise intensity of 0.05. The best ansatz is 3 no-pooling (99.52\%), which achieves almost the same recognition accuracy as the best ansatz under noiseless conditions (99.53\%). The difference in recognition performance between the best and worst (89.83\%) ansatzes is approximately 10\%, which demonstrates the importance of ansatz optimization in practical problems and also indicates that the choice of ansatz is influenced by many real-world factors.

Optimal quantum circuit ansatzes change with different noise types and intensities, and quantum noise can sometimes enhance performance similarly to classical regularization techniques like Dropout. This suggests that specific types of quantum noise might unexpectedly provide regularization benefits on near-term quantum devices.

\subsection{Feature extractor ablation}\label{subsec23}
To evaluate the performance of various feature extractors combined with QCNN and to demonstrate the superiority of the ViT-QCNN-FT results, we designed a series of comparative experiments. The experiments included a 12-qubit QCNN without any feature extractors and a 10-qubit ViT-QCNN without fine-tuning. We refer to the ViT-QCNN without fine-tuning as ViT-QCNN-Base to distinguish it from the fine-tuned model ViT-QCNN-FT. Other experiments involved replacing the fine-tuned ViT in the ViT-QCNN-FT framework with PCA, DCT, Autoencoder, fine-tuned ResNet, fine-tuned EfficientNet, and fine-tuned GoogLeNet. Each method was tested under the 18 QCNN ansatzes.

\begin{figure}[h]
\centering
\includegraphics[width=1\linewidth]{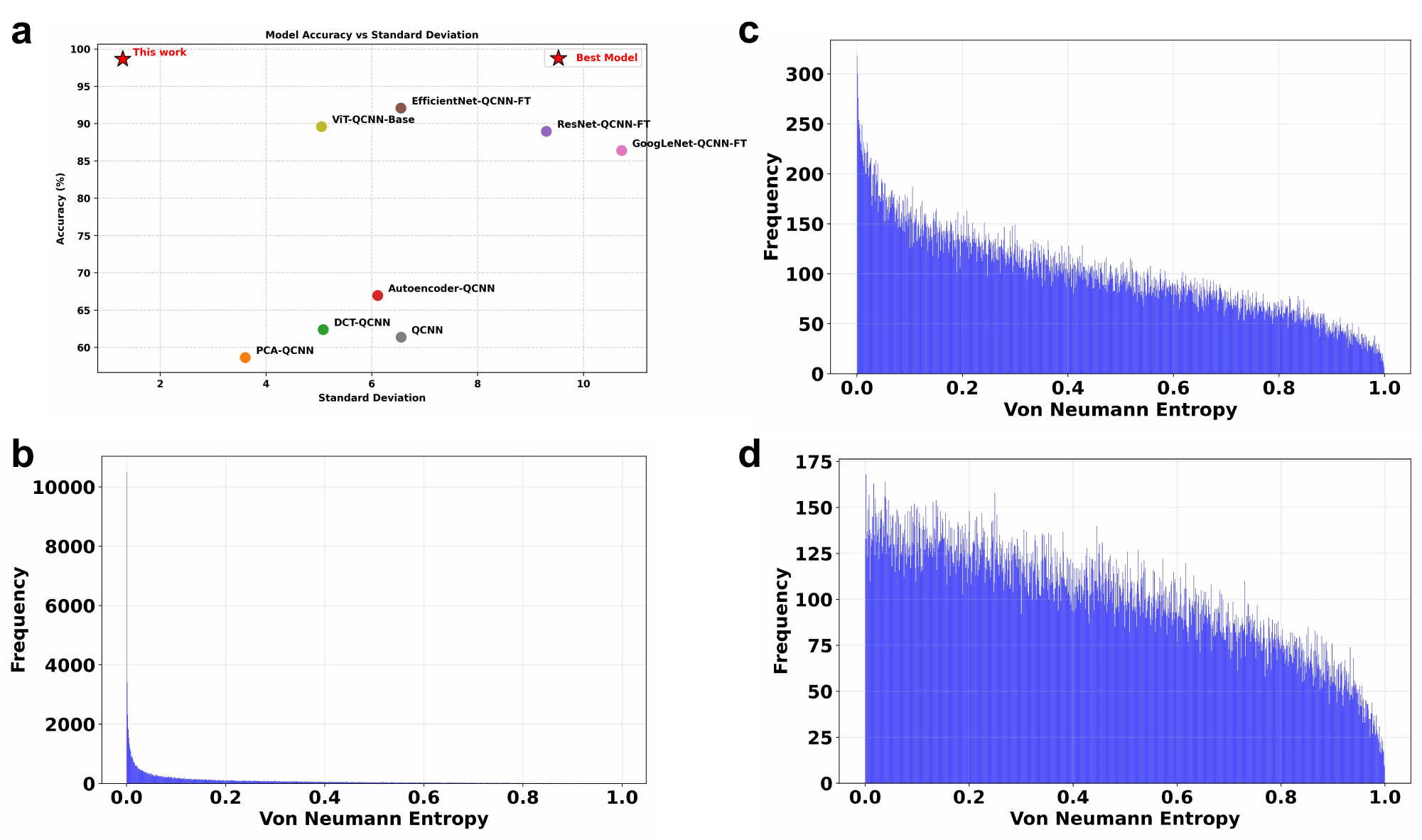}
\caption{\label{FeatureExtract}\textbf{a}, This work is compared with eight other models. The horizontal axis represents the standard deviation, and the vertical axis represents the accuracy. Each point is derived from the average result of the 18 ansatzes under this model. \textbf{b}, The sampling entanglement entropy distribution of convolution 7. \textbf{c}, The sampling entanglement entropy distribution of convolution 8. \textbf{d}, The sampling entanglement entropy distribution of convolution 9. }
\end{figure}

For QCNN and ViT-QCNN-Base, we need to clarify why the 8-qubit is not used. When utilizing QCNN, encoding a $32\times32\times3$ image into a quantum state through amplitude encoding requires a minimum of 12 qubits. In the case of the ViT-QCNN-Base, the pre-trained ViT has a fixed architecture. After removing the classification head, the 768-dimensional output vector serves as the classical input dimension for quantum state encoding, necessitating at least 10 qubits under amplitude encoding. 

Comparative results reveal significant performance discrepancies among these methods (Fig.\ref{FeatureExtract}\textbf{a}). Unsupervised non-deep approaches (PCA/DCT) achieve maximum accuracy of below 65\% through mathematical transformations, while the unsupervised deep learning method (Autoencoder) shows only a marginal improvement, remaining under 70\%. In contrast, supervised deep learning techniques (fine-tuned ResNet, fine-tuned EfficientNet, fine-tuned GoogLeNet) that employ pre-trained networks for feature extraction display substantial gains. The pure 12-qubit QCNN without classical feature compression only achieves 68.04\% as the highest accuracy, performing similarly to PCA\slash DCT\slash Autoencoder. Meanwhile, the 10-qubit ViT-QCNN-Base achieves an impressive 95.09\% accuracy, matching the performance level of 8-qubit ResNet\slash EfficientNet\slash GoogLeNet hybrids. Notably, both higher-qubit approaches exhibit significant gaps in performance compared to the 8-qubit ViT-QCNN-FT. The complete results of the 18 ansatzes for each model are presented in Supplementary Tables 6 and 7 (Supplementary Information). Furthermore, ansatz 8 generally outperforms ansatz 7, despite having the same number of parameters and similar architectures. The relatively more uniform entanglement entropy distribution may contribute to this performance difference. The specific optimal ansatz and accuracy for each method are shown in Table~\ref{table6}. Ansatzes 8 and 9 are usually the best ansatzes. The noise resilience of ansatzes 8 and 9 is generally stronger. We can also observe that the entanglement distributions of their convolution ansatzes are more uniform (Fig.\ref{FeatureExtract}).

\begin{table}[h]
\caption{The optimal ansatzes and accuracy of different methods in 8-qubit amplitude encoding. \#Qubits represents the number of qubits used.}\label{table6}%
\begin{tabular}{@{}llll@{}}
\toprule
\textbf{Methods} & \textbf{\#Qubits} & \textbf{Optimal ansatz} & \textbf{Accuracy} \\
\midrule
QCNN & 12 & ansatz 8 no-pooling & 68.04\%  \\
ViT-QCNN-Base & 10 & ansatz 9 no-pooling & 95.09\% \\
PCA-QCNN & 8 & ansatz 9 no-pooling & 62.13\%  \\
DCT-QCNN & 8 & ansatz 9 no-pooling & 67.64\%  \\
Autoencoder-QCNN & 8 & ansatz 9 pooling & 75.70\%  \\
ResNet-QCNN-FT & 8 & ansatz 4 pooling & 96.11\% \\
EfficientNet-QCNN-FT & 8 & ansatz 8 pooling & 95.27\%  \\
GoogLeNet-QCNN-FT & 8 & ansatz 3 pooling & 94.19\% \\
\botrule
\end{tabular}
\label{table6}
\end{table}

\subsection{Quantum efficiency}\label{subsec24}
To demonstrate the efficiency of QCNN, we replaced the QCNN in 8-qubit amplitude-encoded ViT-QCNN-FT with CNNs that have the same number of parameters (12) or even more (39). 

\begin{figure}[H]
\centering
\includegraphics[width=1\linewidth]{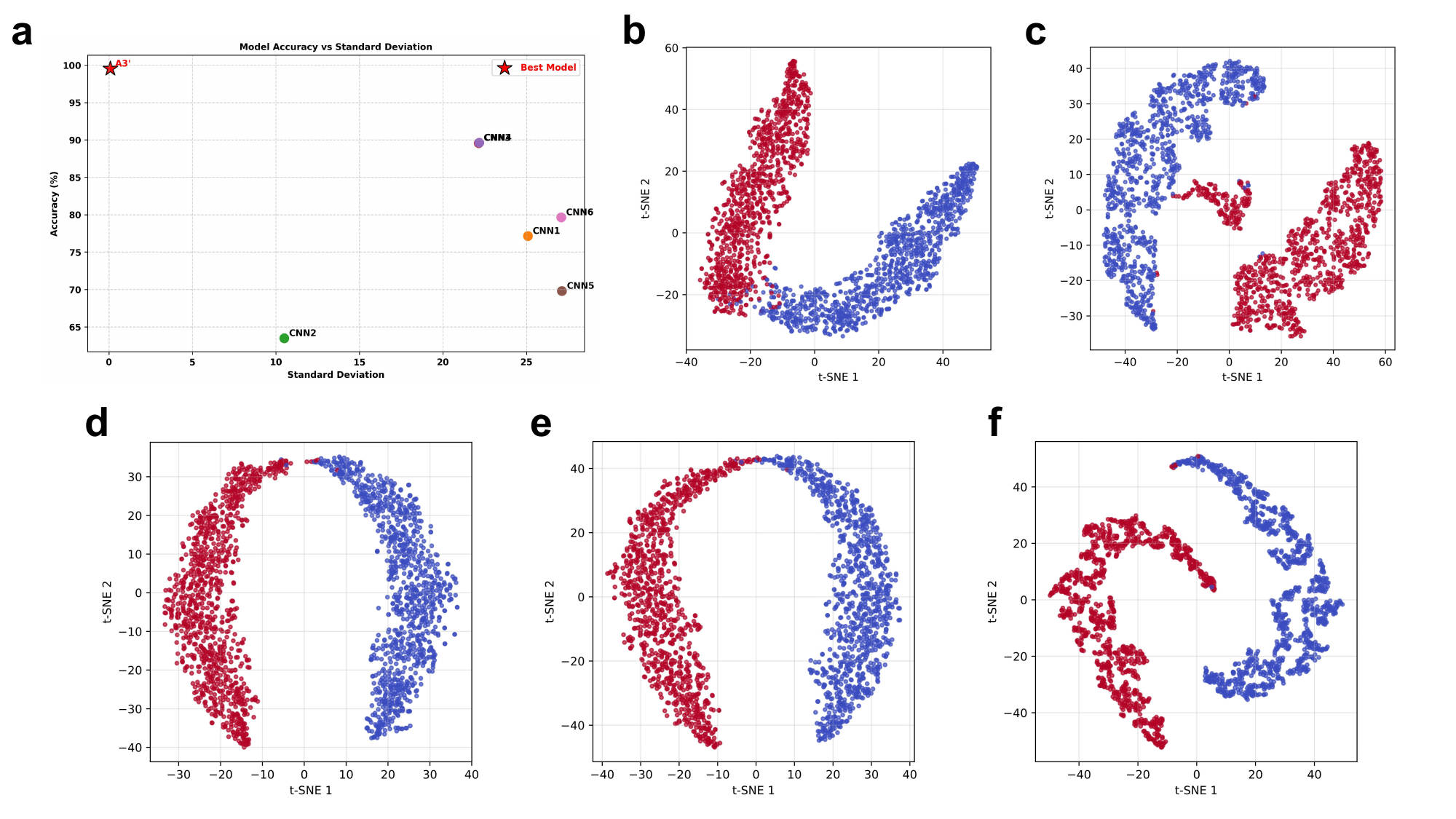}
\caption{\label{QCNN_CNN_compare}\textbf{a}, Comparison between Ansatz 3 no-pooling(A3’) and other classic CNNs. The horizontal axis represents the standard deviation, and the vertical axis represents the accuracy. \textbf{b}, T-SNE of CNN1. \textbf{c}, T-SNE of CNN2. \textbf{d}, T-SNE after the first layer of convolution and pooling in A3'. \textbf{e}, T-SNE after the second layer of convolution and pooling in A3'. \textbf{f}, T-SNE after the third layer of convolution and pooling in A3'. }
\end{figure}

The comparison results are shown in Fig.\ref{QCNN_CNN_compare}\textbf{a}. Here, A3' represents ansatz 3 no-pooling, with 12 parameters. For the classical baselines, CNN1 and CNN2 also consist of 12 parameters each, while CNN3 to CNN6 contain 39 parameters. Despite its relatively small number of parameters, the QCNN achieves superior accuracy and exhibits greater stability compared to the CNN counterparts. 

To further examine the feature representations learned by the models, t-SNE \cite{maaten2008visualizing} was employed to project high-dimensional embeddings into two dimensions for visualization of class separability. Layer-wise t-SNE analyses were performed for A3', CNN1, and CNN2, as illustrated in Fig.\ref{QCNN_CNN_compare}. For A3', the first layer visualizes the remaining 1st, 3rd, 5th, and 7th qubits after convolution and pooling; in the second layer, the 1st and 5th qubits are visualized; and in the third layer, the 5th qubit, which is used for classification, is shown. T-SNE visualization was not performed for the second layer of CNN1 and CNN2 because their output dimension is only 2, rendering t-SNE unsuitable.

The layer-wise t-SNE visualization demonstrates that the QCNN effectively separates and clusters the two classes across all three layers, indicating highly discriminative feature learning. In contrast, the first convolutional layer of CNN1 shows substantial overlap between the two classes, whereas CNN2 exhibits a small cluster of points bridging the classes, preventing a complete separation. These observations suggest that, at the feature level captured by the first layer, the QCNN generates more discriminative embeddings compared to the classical CNN architectures. Notably, QCNNs exploit quantum entanglement to achieve non-classical, high-dimensional feature representations, surpassing the representational capacity of their classical counterparts.

Overall, these results highlight not only the role of quantum entanglement in enhancing classification performance but also the effectiveness of QCNNs within the ViT-QCNN-FT framework.

\section{Methods}\label{sec11}
In this section, we present a comprehensive technical elucidation of the ViT-QCNN-FT algorithm. The overall algorithm is illustrated in Fig.\ref{ViT-QCNN-FT}. The pre-trained ViT is fine-tuned to serve as a feature extractor. The extracted classical features are then encoded into quantum states, and the classification is subsequently carried out by a QCNN.

\subsection{Feature extraction} 
For a pre-trained ViT trained on the dataset $\mathcal{D}_{p} \subset \mathbb{R}^{L_1\times W_1 \times C_1}$, where $L_1$,  $W_1$, and $C_1$ represent the length, width and channels of its images, it can be expressed as $ V_{p}:\mathbb{R}^{L_1\times W_1 \times C_1} \to \mathbb{R}^{k} $, where $k$ represents the number of classes for $\mathcal{D}_{p}$. Let $M_{p}$ be the multilayer perceptron (MLP) of $V_{p}$, referred to as the classification head, and $E_{p}$ represent other components of $V_{p}$, excluding the final classification head. Thus, we have $V_{p}=M_{p} \circ E_{p}$.

Due to the differences between the target dataset $\mathcal{D}_{t} \subset \mathbb{R}^{L_2 \times W_2 \times C_2}$ and $\mathcal{D}_{p}$, where $L_2$,  $W_2$, and $C_2$ represent the length, width and channels of its images, it is necessary to fine-tune $V_{p}$ on $\mathcal{D}_{t}$. First, we must perform a simple preprocessing on $\mathcal{D}_{t}$ to ensure it matches the input shape required by $V_{p}$. By replacing the MLP $M_{p}$ with a new MLP $M_{t}$, we get $V_{t}=M_{t} \circ E_{p}$, where $V_{t}:\mathbb{R}^{L_1\times W_1 \times C_1} \to \mathbb{R}^m$, with $m$ indicating the number of classes for $\mathcal{D}_{t}$. Note that the parameters of $M_{t}$ are randomly initialized. We then use a subset of $\mathcal{D}_{t}$ to train $V_{t}$. During this training process, the parameters of $E_{p}$ remain fixed, while only the parameters of $M_{t}$ are updated. 

After training, we truncate $M_{t}$ to retain only the initial layers, which are combined with $E_{p}$ to create a feature extractor. In other words, if $M_{t}=M_{d} \circ M_{s}$, the feature extractor can be represented as $V_{e}=M_{s} \circ E_{p}$ with $V_{e}:\mathbb{R}^{L_1\times W_1 \times C_1} \to \mathbb{R}^N$, where $N$ signifies the output dimensionality. We then apply the feature extractor $V_{e}$ to the remaining data in $\mathcal{D}_{t}$, resulting in a dataset $\mathcal{D}$ that contains the extracted features. 

\subsection{Quantum state encoding}
In quantum machine learning, mapping classical data to quantum states is a crucial step in achieving quantum advantage. For simplicity, let us consider the dataset $\mathcal{D}=\{\mathbf{x}_i\}_{i=1}^M,$ where $\mathbf{x}_i \in \mathbb{R}^N$, $N$ denotes the dimension of the data. These classical data are mapped to a quantum state $\ket{\psi(\mathbf{x})}$, which belongs to a Hilbert space $\mathcal{H}$. This process is known as quantum state encoding (green block in Fig.\ref{ViT-QCNN-FT}), which may also be referred to as data embedding, data upload, or data encoding. Below, we will introduce several methods for quantum state encoding. 

\subsubsection{Amplitude Encoding}
Amplitude encoding is one of the most widely used methods for quantum state representation \cite{caro2021encoding}. It represents data $\mathbf{x}=(x_1, \cdots, x_N)^T$ of dimension $N = 2^n$ as the amplitudes of an n-qubit quantum state. Specifically, the quantum state $\ket{\psi(\mathbf{x})}$ is defined as:
\begin{equation}
\label{eq:amplitude}
\ket{\psi(\mathbf{x})} = \frac{1}{\|\mathbf{x}\|} \sum_{i=1}^{N} x_i \ket{i}, 
\end{equation}
where $\ket{i}$ is the $i$-th computational basis state, and $\|\mathbf{x}\|$ denotes the Euclidean norm (or $L_2$-norm) of the vector $\mathbf{x}$, ensuring the normalization of the quantum state. Amplitude encoding provides an efficient means to represent classical data in quantum systems, as it allows $N$ classical data to be encoded into $log(N)$ qubits, significantly reducing the number of qubits required to represent high-dimensional data.

\subsubsection{Angle encoding}
Angle encoding encodes data features into the rotation angles of parameterized quantum gates acting on qubits \cite{mitarai2018quantum}. It embeds one classical data point $x_i$, which is scaled to the range between $0$ and $\pi$ , into a single qubit as $\ket{\phi(x_i)} = \cos(\frac{x_i}{2})\ket{0} + \sin(\frac{x_i}{2})\ket{1}$ for $i = 1, \cdots, N$. Therefore, angle encoding transforms $\mathbf{x}=(x_1, \cdots, x_N)^T$ into N qubits as 
\begin{equation}
\label{eq:qubit}
    \ket{\psi(\mathbf{x})}  = \bigotimes_{i=1}^{N} (\cos(\frac{x_i}{2})\ket{0} + \sin(\frac{x_i}{2})\ket{1}),
\end{equation}
where $x_i\in \lbrack 0,\pi)$ for all $i$. 
This can be achieved using the gate $R_y(\theta)$, that is:
\begin{equation}
\label{eq:angle}
    \ket{\psi(\mathbf{x})}  = \bigotimes_{i=1}^{N} R_y(x_i)\ket{0}.
\end{equation}
Setting $N$ initial qubits to $\ket{0}$, each qubit undergoes the corresponding rotation gate $R_y(x_i)$, resulting in the system state being the angle-encoded state $\psi(\mathbf{x})$. 

\subsubsection{Dense angle encoding}
The angle encoding mentioned above can be generalized to encode two classical data points into a single qubit by using rotations around two orthogonal axes \cite{larose2020robust}. Choosing them to be the x and y axes of the Bloch sphere, dense angle encoding encodes $\mathbf{x}_k=(x_{k_1}, x_{k_2})$ as 
\begin{equation}
    \ket{\phi(\mathbf{x}_k)} = e^{-i\frac{x_{k_2}}{2}\sigma_{y}}e^{-i\frac{x_{k_1}}{2}\sigma_{x}} \ket{0}.
\end{equation}
Therefore, the dense angle encoding maps $\mathbf{x}=(x_1, \cdots, x_N)^T$ to $\frac{N}{2}$ qubits as 
\begin{equation}
\label{eq:dense}
    \ket{\phi(\mathbf{x})}  = \bigotimes_{j=1}^{N/2} \left( e^{-i\frac{x_{N/2+j}}{2}\sigma_{y}}e^{-i\frac{x_{j}}{2}\sigma_{x}} \ket{0} \right).
\end{equation}
Classical data can be arbitrarily paired and encoded into individual qubits. 

\subsection{Quantum convolution neural network} 
QCNN is a quantum machine learning model introduced in recent years, inspired by classical neural networks. Theoretical analyses suggest that QCNN, due to its local architecture design and hierarchical information processing mechanism, can avoid the issue of barren plateaus (i.e., the exponential decay of gradients with system size) \cite{pesah2021absence}. In this section, we will provide a detailed introduction to QCNN, demonstrate its trainability, and explain how to optimize its parameters after measuring the quantum circuit. 

\subsubsection{Convolution ansatzes and pooling ansatz}
The specific QCNN illustrated in Fig.\ref{ViT-QCNN-FT} consists of three layers. Each layer includes two components: the convolutional layer and the pooling layer. The convolutional layer is a core element of the QCNN, comprising parameterized quantum circuit modules that operate on adjacent pairs of qubits in a translationally invariant manner. This means that the quantum circuit modules within the same convolutional layer are identical. The action of the two-qubit parameterized quantum circuit $U_c$ on the two-qubit density matrix $\rho$ can be expressed as:
\begin{equation}
    \rho'= U_c\rho U^\dagger_c.
\end{equation}
The convolutional layer is responsible for extracting local features from the input quantum state, while the pooling layers reduce the size of the quantum system: 
\begin{equation}
    \rho_B=\text{Tr}_A \left(U_p \rho_{AB} U^{\dagger}_{p}\right),
\end{equation}
where $\Tr_A\left( \cdot \right)$ denotes a partial trace over subsystem $A$, $\rho_{AB}$ is a two-qubit state to be pooled, and $U_{p}$ is the unitary operation represented by the pooling ansatz. In QCNN, the number of parameters in both the convolutional layer and pooling layer is independent of the system size, significantly reducing the number of parameters that need to be optimized.

The convolution ansatzes and pooling ansatz examined in this study are depicted in Fig.\ref{ViT-QCNN-FT}. These circuits have been tested in a previous study \cite{hur2022quantum}. Most of these ansatzes draw inspiration from prior research. For example, circuit (a) is used as the parameterized quantum circuit for training a tree tensor network~\cite{grant2018hierarchical}. Circuits (b), (c), (d), (e), (g), and (h) are based on the work of Sim et al.\cite{sim2019expressibility}, which presents an analysis of the expressibility and entangling capacity of four-qubit parameterized quantum circuits. These have been adapted into two-qubit versions to form the fundamental components of the convolutional layer. Specifically, circuits (g) and (h) are simplified versions of those circuits that showed the highest expressibility in Sim et al.'s analysis. Circuit (b) is a two-qubit variant of the quantum circuit that demonstrated the strongest entangling capability. Circuits (c), (d), and (e) are chosen for their balanced combination of expressibility and entangling power. Circuit (f) was designed as an appropriate candidate for a two-body entangler within the variational quantum eigensolver (VQE) framework \cite{parrish2019quantum}. This circuit is also recognized for its ability to implement arbitrary $SO(4)$ gates~\cite{wei2012decomposition}. Lastly, circuit (i) corresponds to the parameterization of an arbitrary $SU(4)$ gate~\cite{vatan2004optimal,maccormack2022branching}.

These ansatzes modulate the intensity of entanglement through various two-qubit gates (e.g., CNOT\slash CRX\slash CRZ), which directly influence the feature extraction capabilities. The entanglement level is always quantified using the von Neumann entropy, defined as:
\begin{equation}
S_{\text{VN}}(\rho_{AB}) = -\text{Tr}(\rho_A \log_2 \rho_A), \quad \rho_A = \text{Tr}_B(\rho_{AB}).
\end{equation}
We randomly initialize the parameters of each convolution ansatz 100,000 times. The distributions of von Neumann entropies are illustrated in Supplementary Figures 2 and 3 (Supplementary Information). The entanglement entropies for all ansatzes lie within the range 0 to 1, indicating a moderate level of entanglement. However, their distribution patterns exhibit some distinct differences. The entanglement entropy distribution of ansatz 2 is omitted, as its value remains fixed at 1. As shown in Fig.\ref{ViT-QCNN-FT}, convolutions 4 and 5 exhibit similar structures, as do convolutions 7 and 8. The entanglement distributions of convolutions 4 and 5 are nearly identical, while convolution 8 displays a more balanced distribution compared to convolution 7, whose entropy is more concentrated at lower values; this corresponds to its stronger noise robustness observed in experiments. Convolution 9 exhibits the most uniformly balanced distribution and frequently emerges as the optimal ansatz in our experiments.

\subsubsection{Quantum convolution neural network ansatzes}\label{sec332}
A QCNN ansatz can be constructed by combining a convolution ansatz with a pooling ansatz, resulting in a PQC within the ViT-QCNN-FT framework that requires optimization. The configurations of the 18 QCNN ansatzes employed in our simulations are summarized in Table~\ref{table1}, while the corresponding numbers of trainable parameters for an 8-qubit quantum circuit are reported in Table~\ref{table2}.

\begin{table}[h]
\caption{The configurations of the 18 QCNN ansatzes. They are divided into two parts: pooling and no-pooling. $\{\text{C}i,\text{P}\}^9_{i=1}$ indicates the convolutional layer is composed of `Convolution $i$' (Fig.\ref{ViT-QCNN-FT}) and the pooling layer is composed of `Pooling' (Fig.\ref{ViT-QCNN-FT}). $\{\text{C}i,\text{-}\}^9_{i=1}$ indicates the convolutional layer is composed of `Convolution $i$' and the pooling layer is composed of partial trace operation.}\label{table1}%
\begin{tabular}{@{}lccccccccc@{}}
\toprule
\textbf{Ansatz} & \textbf{1} & \textbf{2} & \textbf{3} & \textbf{4} & \textbf{5} & \textbf{6} & \textbf{7} & \textbf{8} & \textbf{9} \\
\midrule
pooling & C1, P & C2, P & C3, P & C4, P & C5, P & C6, P & C7, P & C8, P & C9, P \\
no-pooling & C1, - & C2, - & C3, - & C4, - & C5, - & C6, - & C7, - & C8, - & C9, - \\
\botrule
\end{tabular}
\label{table1}
\end{table}

By examining the circuit architecture of the QCNN, we observe that quantum information progressively converges toward the classification qubit as the layers deepen. We randomly initialized the QCNN parameters 100,000 times, and the classification qubit entanglement distributions for the 18 QCNN ansatzes are provided in Supplementary Figures 4-7 (Supplementary Information). Our analysis indicates that the stacking of convolutional and pooling layers enhances global qubit entanglement, thereby improving the quality of feature representations. Taking anstaz 8 no-pooling as an example, the average entanglement entropy of the classified qubits is approximately 0.61 after the first layer, 0.87 after the second layer, and 0.91 after the third layer.

\begin{table}[ht]
\centering
\caption{Number of parameters of the 18 ansatzes in 8-qubit QCNN.}
\label{table2}
\begin{tabular}{lccccccccc}
\toprule
\textbf{Ansatz} & \textbf{1} & \textbf{2} & \textbf{3} & \textbf{4} & \textbf{5} & \textbf{6} & \textbf{7} & \textbf{8} & \textbf{9} \\
\midrule
Pooling     & 12 & 12 & 18 & 24 & 24 & 24 & 36 & 36 & 51 \\
No-pooling  &  6 &  6 & 12 & 18 & 18 & 18 & 30 & 30 & 45 \\
\bottomrule
\end{tabular}
\end{table}

\subsection{Trainability}
A significant challenge in training PQCs is the phenomenon of barren plateaus, where the gradients of the cost function vanish exponentially as the number of qubits or the circuit depth increases \cite{mcclean2018barren}. This vanishing gradient problem prevents gradient-based optimizers from effectively updating parameters, leading to training failure. 

Research indicates that barren plateaus typically arise in deep or highly entangled PQCs employing global cost functions, as the unitary transformations implemented by such circuits form approximate 2-designs, causing the gradient variance to decay as $O(1/2^n)$, where $n$ is the number of qubits \cite{cerezo2021cost}. The variance of the cost function gradient does not depend on the size of the entire quantum system, but only on the number of qubits within the causal cone of the measurement observable. Suppose our cost function $L$ is defined as the expectation value of a local observable $O_v$ acting on one or a few qubits $v$:
\begin{equation}
    L = \langle \psi(\boldsymbol{\theta}) | O_v | \psi(\boldsymbol{\theta}) \rangle
\end{equation}
where $|\psi(\boldsymbol{\theta})\rangle = U(\boldsymbol{\theta})|0\rangle^{\otimes n}$ is the quantum state prepared by the PQC with parameters $\boldsymbol{\theta}$. The variance $\text{Var}[\partial_k L]$ of the partial derivative gradient $\partial_k L$ of the cost function with respect to parameter $\theta_k$ becomes a key metric for assessing trainability. Exponentially small variance indicates the presence of barren plateaus. 

For QCNNs, due to the geometric locality of their convolutional and pooling operations, the causal cone $C(O_v)$ of a local observable $O_v$ acting on a small number of output qubits $v$ does not scale exponentially with the total number of qubits $n$. Instead, the hierarchical structure of the QCNN ensures that the size of the causal cone grows very slowly, typically logarithmically, i.e., $|C(O_v)| \sim O(\log n)$. For such architectures with local cost functions, the lower bound on the gradient variance scales inversely polynomially with the number of qubits $|C(O_v)|$ within the causal cone. Specifically, the gradient variance satisfies the inequality \cite{pesah2021absence}:
\begin{equation}
    \text{Var}[\partial_k L] \geq \frac{F_1}{\text{poly}(|C(O_v)|)}
\end{equation}
where $F_1$ is a constant dependent on the specific gates used, and $\text{poly}(\cdot)$ is a polynomial function. This expression is central to avoiding barren plateaus. It shows that as long as the size of the causal cone $|C(O_v)|$ grows slowly (logarithmically in the case of QCNNs), the gradient variance does not vanish exponentially with increasing total qubit number $n$. Furthermore, they provided a more concrete lower bound on the gradient norm to demonstrate trainability \cite{pesah2021absence}:
\begin{equation}
    \sum_{k} (\partial_k L)^2 \geq \frac{F_2}{q(D, w)}
\end{equation}
where $F_2$ is a constant, $D$ is the circuit depth, $w$ is a width parameter related to the circuit structure, and $q(\cdot)$ is a polynomial. This expression guarantees the existence of at least one direction where the gradient does not vanish too rapidly, thereby ensuring the existence of a viable optimization path.

In summary, QCNNs, through the synergistic combination of their hierarchical architecture and local cost functions, effectively confine gradient calculations to a logarithmically scaling causal cone. This fundamentally breaks the conditions leading to barren plateaus, guaranteeing the model's trainability and establishing QCNNs as a highly promising quantum machine learning model with significant potential for scaling to large quantum systems \cite{cong2019quantum, pesah2021absence}.

\subsection{Learning process}

Having established the fusion framework integrating the fine-tuned ViT feature extractor with quantum state encoding and QCNN architecture, we now formalize the critical parameter optimization procedure. In our quantum binary classification scheme, for the $i$-th input sample, the parameterized quantum circuit prepares a final state $\ket{\psi^{(i)}(\bm{\theta})}$, where $\bm{\theta} = \left[\theta_1, \dots, \theta_d \right]^T$ denotes the trainable parameters. The classification is performed by measuring the readout qubit in the computational basis, yielding the probability distribution:
\begin{equation}
\bm{p}^{(i)}(\bm{\theta}) = \left[p^{(i)}_0(\bm{\theta}),\  p^{(i)}_1(\bm{\theta})\right]^T, 
\end{equation}
where $p^{(i)}_0(\bm{\theta}) = |\langle 0|\psi^{(i)}(\bm{\theta})\rangle|^2$ and $p^{(i)}_1(\bm{\theta}) = |\langle 1|\psi^{(i)}(\bm{\theta})\rangle|^2$. 
The probability for class $y=1$ is directly given by the second component:
\begin{equation}
p^{(i)}(\bm{\theta}) = p^{(i)}_1(\bm{\theta}). 
\end{equation}
For the true label $y^{(i)} \in \{0,1\}$ of $M$ training samples, we construct the binary cross-entropy cost function:
\begin{equation}
L(\bm{\theta}) = -\frac{1}{M} \sum_{i=1}^{M} \left[ y^{(i)} \log p^{(i)}(\bm{\theta}) + (1 - y^{(i)}) \log(1 - p^{(i)}(\bm{\theta})) \right]. 
\end{equation}
Although gradient-free optimization techniques offer noise resilience for variational quantum circuits \cite{peruzzo2014variational, crooks2018gradient}, we adopt gradient-based optimization due to its superior convergence rate and parameter efficiency in high-dimensional quantum models \cite{schuld2019gradient, stokes2020quantum}. Parameter optimization is performed using gradient descent:
\begin{equation}
\theta_j \leftarrow \theta_j - \eta \frac{\partial L}{\partial \theta_j},
\end{equation}
where \( \eta \) is the learning rate. The gradient is computed using the chain rule: 
\begin{equation}
\frac{\partial L(\bm{\theta})}{\partial \theta_j} = -\frac{1}{M} \sum_{i=1}^{M}   \left(
\frac{\partial L(\bm{\theta})}{\partial p^{(i)}(\bm{\theta})} \cdot 
\frac{\partial p^{(i)}(\bm{\theta})}{\partial \theta_j} \right). 
\end{equation}
The partial derivatives can be calculated as follows:
\begin{align}
\frac{\partial L(\bm{\theta})}{\partial p^{(i)}(\bm{\theta})} &=   \frac{y^{(i)}}{p^{(i)}(\bm{\theta})} - \frac{1 - y^{(i)}}{1 - p^{(i)}(\bm{\theta})},  \\
\frac{\partial p^{(i)}(\bm{\theta})}{\partial \theta_j} &= \frac{ p^{(i)}(\theta_j + \frac{\pi}{2}) - p^{(i)}(\theta_j - \frac{\pi}{2}) }{2}, 
\end{align}
where $p(\theta_j \pm \frac{\pi}{2})$ denotes the measured probability of $|1\rangle$ when only parameter $\theta_j$ is shifted by $\pm\frac{\pi}{2}$ while all other parameters remain fixed \cite{li2017hybrid,mitarai2018quantum}.
Thus, the expression for the gradient becomes:
\begin{equation}
\frac{\partial L}{\partial \theta_j} = 
-\frac{1}{2M} \sum_{i=1}^{M} 
\left( \frac{y^{(i)}}{p^{(i)}(\bm{\theta})} - \frac{1 - y^{(i)}}{1 - p^{(i)}(\bm{\theta})} \right) 
\left( p^{(i)}\left(\theta_j + \frac{\pi}{2}\right) - p^{(i)}\left(\theta_j - \frac{\pi}{2}\right) \right). 
\end{equation}
Parameters should be updated iteratively until either convergence is achieved or specified termination conditions are met.

\section{Discussion}\label{sec12}
This study systematically explores the combined effect of fine-tuned pre-trained ViT with various QCNN ansatzes through comprehensive experimental analysis across encoding methodologies, quantum noise sensitivity, structure design, ansatz effect, entanglement influence, and classical-quantum hybrid model performance. 

The choice of quantum encoding significantly influences model effectiveness. With 10-qubit amplitude encoding, the ViT-QCNN-FT achieves an average accuracy of 98.88\% across ansatzes, starkly contrasting with the 90.09\% observed under angle encoding. This 8.79\% performance gap underscores the importance of efficient classical-to-quantum data transformation and highlights the need for innovations such as data-driven variational quantum encoding.

Interestingly, quantum noise demonstrates unexpected regularization potential. In simulations of amplitude damping noise with intensity 0.01, ansatz 1 pooling improves recognition accuracy by 2.71\%. This phenomenon indicates that noise might help escape local minima during optimization, encouraging further research into intentional noise modulation to enhance performance. 

In terms of QCNN structural design, a comparative evaluation of 18 QCNN ansatzes under ideal and noisy conditions shows that without pooling ansatz exhibit a small impact on overall classification accuracy. For instance, the accuracy difference $\Delta_{\text{acc}}$ between pooling and no-pooling configurations under 8-qubit amplitude encoding satisfies $\Delta_{\text{acc}} < 0.15\%$.  This finding indicates that when designing QCNN, it is unnecessary to adhere strictly to the classical CNN paradigm of "\text{convolution + pooling}". The essential components to retain in QCNN are the translation-invariant design of the convolution layer and the trace operation following that layer. This has significant guiding value for exploring more flexible and resource-efficient quantum circuit ansatzes.

Ansatz is crucial to performance, with a 33.15\% accuracy differential observed under angle encoding. This variability necessitates a future focus on automated ansatz generation, structural search algorithms, and task-adaptive optimization.

Entanglement entropy provides a useful guideline for ansatz design. Moderate entanglement enhances feature extraction by enabling a richer representation space, while constraining entanglement ($0<S_\text{VN}<1$) can act as a natural regularizer under noise. This observation aligns with classical findings that limited model capacity may improve generalization~\cite{vatan2004optimal}. Notably, convolutional ansatzes with uniform and higher average von Neumann entropy, such as convolutions 8 and 9, exhibit stronger robustness to noise. The ablation experiments further indicate that, even with similar structures and parameter counts, uniform entanglement in ansatz 8 leads to superior feature extraction compared to ansatz 7. These results highlight the critical role of entanglement in balancing expressivity and stability in quantum feature extractors.

The synergistic advantage of fine-tuned pre-trained ViT feature extraction is confirmed through ablation studies. Its accuracy surpasses that of alternatives (such as PCA, DCT, Autoencoder, and fine-tuned ResNet\slash EfficientNet\slash GoogLeNet) by 6.64\%–40.56\% in hybrid architectures. Notably, replacing QCNN with classical CNN counterparts while keeping the same parameter counts results in an average accuracy decrease of 29.36\%, demonstrating QCNN's superior capabilities in information compression and nonlinear mapping while affirming the potential of classical-quantum hybridization. 

All experiments conducted in this study were performed in a quantum simulator environment. The complex noise types and gate errors present in real quantum devices have not been fully investigated. Additionally, the current experiments focus on the binary classification task of color images, with the generalization ability for multi-classification and multi-modal tasks requiring further exploration.

Overall, our results demonstrate that classical pre-processing and quantum feature fusion are complementary and mutually reinforcing. The challenges identified in ansatz design and optimization define a clear research agenda, while the proposed hybrid framework provides a scalable template for future work. We anticipate that these principles of classical–quantum hybridization will play a central role in realizing the full potential of quantum machine learning for real-world applications.





\bibliography{sn-bibliography,sn-article.bbl}

\end{document}